\documentclass[reprint,amsmath,amssymb,aps,superscriptaddress,nofootinbib]{revtex4-1}
\usepackage{graphicx}
\usepackage{dcolumn}
\usepackage{bm}
\usepackage[utf8]{inputenc}

\usepackage{hyperref}
\usepackage{amsmath, amsthm, amsfonts,amssymb}
\usepackage[svgnames]{xcolor}
\usepackage{mathtools}
\usepackage{caption}
\usepackage{subcaption}
\usepackage{xcolor}
\usepackage{soul}
\usepackage{xcolor}
\usepackage{appendix}
\usepackage{orcidlink}
\usepackage{booktabs}
\usepackage{rotating}
\usepackage{comment}
\usepackage{dblfloatfix}

\usepackage{booktabs}
\usepackage{siunitx}
\definecolor{teal}{RGB}{0, 128, 128}
\hypersetup{ colorlinks=true, linkcolor=teal, citecolor=teal, urlcolor=teal}

\begin{document}

\title{Chiral symmetry and curvature bounds in de Sitter spacetime}

\author{L. G. Barbosa \orcidlink{0009-0007-3468-3718}}
\email{leonardo.barbosa@posgrad.ufsc.br}
\affiliation{Departamento de Física, CFM - Universidade Federal de \\ Santa Catarina; C.P. 476, CEP 88.040-900, Florianópolis, SC, Brazil}

\begin{abstract}
We study chiral symmetry breaking in the Nambu--Jona-Lasinio model on a de Sitter background, treating it as a non-renormalizable effective field theory with a physical ultraviolet cutoff. Using exponential proper-time regularization, we obtain an exact solution for the constituent fermion mass in the strong-curvature regime via the Lambert~$W$ function. The consistency condition for real-valued solutions leads to an upper bound on the cosmological constant, indicating a limitation of the mean-field description rather than a fundamental physical constraint on the spacetime geometry.
\end{abstract}

\maketitle

\section{Introduction}
\label{sec:introduction}

The Nambu--Jona-Lasinio (NJL) model provides a phenomenological framework for studying dynamical chiral symmetry breaking and fermion mass generation~\cite{Nambu:1961tp,Nambu:1961fr,Inagaki:1993ya,Elizalde:1993kb,Inagaki:1997kz,Hill:1991jc,Addazi:2017qus}. When extended to curved backgrounds such as de Sitter spacetime, the interplay between geometric curvature and the fermionic condensate becomes relevant for understanding possible modifications of the vacuum structure in the early universe~\cite{Inagaki:1995jp,Kanemura:1994rs,Kanemura:1995sx,Ishikawa:1996yx}.

Previous analyses of the NJL model in curved spaces often employ subtraction schemes to handle ultraviolet divergences, focusing on the conditions under which curvature may restore chiral symmetry~\cite{Elizalde:1994fh,Geyer:1996wg,Geyer:1996np}. In those studies, the spacetime curvature is typically treated as an independent external parameter, and the validity of the model is assumed over a wide range of curvature scales.

The present work adopts a different perspective by treating the NJL model explicitly as a non-renormalizable effective field theory (EFT), in which the ultraviolet cutoff $\Lambda_{\mathrm{UV}}$ is retained as a physical scale. Using an exponential proper-time regularization scheme~\cite{Saharian:2021mzf}, we examine how the inclusion of this cutoff links the geometric expansion rate to the energy limits of the effective description. The key novel ingredient is the identification of the regularization parameter with the physical cutoff, which enables an exact analytical solution of the gap equation in the strong-curvature regime. Rather than focusing primarily on the phase transition threshold, we investigate the mathematical consistency of the gap equation solution in this regime.

Within this EFT approach, the gap equation is solved analytically in terms of the Lambert $W$ function. The consistency conditions for real-valued solutions then lead to a relation of the form
\begin{equation}
    \Lambda^{\max} \propto \Lambda_{\mathrm{UV}}^{2}.
\end{equation}
This result constitutes a limitation of the model: for a given cutoff $\Lambda_{\mathrm{UV}}$, there exists a maximum curvature beyond which the mean-field gap equation no longer admits a real solution for the constituent mass. Such an outcome is a direct consequence of the specific regularization and of the treatment of the NJL model as an EFT.

The remainder of this paper is organized as follows. Section~\ref{sec:Nambu_Jona_Lasinio_model_in_curved_space_time} reviews the NJL model in curved spacetime and the mean-field approximation, summarizing established results from the literature. Section~\ref{sec:condensate_desitter} reviews the evaluation of the fermionic condensate in de Sitter space using the exponential regularization scheme, following the approach of Ref.~\cite{Saharian:2021mzf}. Section~\ref{sec:UV_expansion} then performs the ultraviolet expansion of the condensate within the effective field theory perspective. Finally, Section~\ref{sec:gap_equation} analyzes the gap equation and the resulting constraints on chiral symmetry, leading to the conclusions in Section~\ref{Conclusions}.

\section{Nambu--Jona-Lasinio model in curved space-time}
\label{sec:Nambu_Jona_Lasinio_model_in_curved_space_time}

In this section, we review the NJL model minimally coupled to a gravitational background~\cite{Inagaki:1993ya}.
The Lagrangian density in curved spacetime reads
\begin{equation}
    \mathcal{L}_{\mathrm{NJL}} = \overline{\psi}\bigl(i\gamma^{\mu}(x)\nabla_{\mu} - m\bigr)\psi
    + G\bigl[(\overline{\psi}\psi)^{2} + (\overline{\psi}i\gamma_{5}\psi)^{2}\bigr],
\end{equation}
where $\psi$ is the Dirac fermion field, $\overline{\psi}$ is its adjoint, $m$ is the current fermion mass, and $G$ is the scalar four-fermion coupling constant. The coordinate-dependent Dirac matrices are $\gamma^{\mu}(x)=e_{a}^{\mu}\gamma^{a}$, with $\gamma^{a}$ the usual flat-space matrices, and the tetrads $e_{a}^{\mu}$ satisfy $g_{\mu\nu}=e_{\mu}^{a}e_{\nu}^{b}\eta_{ab}$. The covariant derivative $\nabla_{\mu}=\partial_{\mu}+\Gamma_{\mu}$ contains the spin connection $\Gamma_{\mu}$, whose standard expression in terms of the tetrads and the affine connection can be found in~\cite{Inagaki:1993ya} and references therein.

In the mean-field approximation the scalar four-fermion term is linearized as
\begin{equation}
    (\overline{\psi}\psi)^{2}\simeq 2\langle\overline{\psi}\psi\rangle\,\overline{\psi}\psi - \langle\overline{\psi}\psi\rangle^{2},
\end{equation}
while the pseudo-scalar condensate is taken to vanish, $\langle\overline{\psi}i\gamma_{5}\psi\rangle=0$, since $\bar{\psi}i\gamma_5\psi$ is a pseudoscalar and the Bunch-Davies 
vacuum preserves parity symmetry. The effective Lagrangian then becomes
\begin{equation}
    \mathcal{L}_{\mathrm{NJL}}^{\mathrm{MF}} = \overline{\psi}\bigl(i\gamma^{\mu}\nabla_{\mu} - M\bigr)\psi - G\langle\overline{\psi}\psi\rangle^{2},
\end{equation}
with the constituent fermion mass $M$ determined self-consistently by the gap equation
\begin{equation}
    M = m - 2G\langle\overline{\psi}\psi\rangle.
\end{equation}

This framework, equivalent to minimizing the effective potential, is well suited for investigating dynamical chiral symmetry breaking in cosmological settings.

\section{Fermionic condensate in de Sitter spacetime}
\label{sec:condensate_desitter}

In this section, we review the evaluation of the fermionic condensate in a de Sitter background, following the exponential regularization scheme established in \cite{Saharian:2021mzf}. We consider the flat Friedmann-Lemaître-Robertson-Walker (FLRW) patch of the de Sitter metric, with the line element given by
\begin{equation}
    ds^{2}=dt^{2}-e^{2t/\alpha}\sum_{i=1}^{3}(dz^{i})^{2},
\end{equation}
where $x^{\mu}=(t,z^{1},z^{2},z^{3})$ are the comoving coordinates and the scale factor is governed by $\alpha = \sqrt{3/\Lambda}$.

The dynamics of the fermion field $\psi$ is dictated by the Dirac equation in curved spacetime:
\begin{equation}
    i\gamma^{\mu}(x)\left(\partial_{\mu}+\Gamma_{\mu}\right)\psi - M\psi = 0,
\end{equation}
where $M$ is the constituent mass determined self-consistently by the NJL gap equation. Given a complete set of mode solutions $\{\psi_{\beta}^{(+)}, \psi_{\beta}^{(-)}\}$ corresponding to the Bunch-Davies vacuum, the fermionic condensate is defined as the vacuum expectation value:
\begin{equation}
    \langle \overline{\psi}\psi \rangle = \frac{1}{2}\sum_{\beta} \left( \overline{\psi}_{\beta}^{(-)}\psi_{\beta}^{(-)} - \overline{\psi}_{\beta}^{(+)}\psi_{\beta}^{(+)} \right),
\end{equation}
where the sum over $\beta$ represents the integration over the continuous and discrete quantum numbers. In the de Sitter manifold, these modes are expressed in terms of Hankel functions, allowing the condensate to be written in the following integral form:
\begin{equation}
    \langle \overline{\psi}\psi \rangle = \frac{8\alpha^{-3}}{(2\pi)^{5/2}} \int_{0}^{\infty} dx \, x^{3/2} e^{x} \text{Im} \left[ K_{1/2-iM\alpha}(x) \right],
\end{equation}
where $K_{\nu}(x)$ is the modified Bessel function of the second kind (Macdonald function).

The integral in its bare form is ultraviolet (UV) divergent. To isolate the physical contributions within the EFT framework, we implement an exponential cutoff regularization, introducing the parameter $s$:
\begin{equation}
    \langle \overline{\psi}\psi \rangle^{(s)} = \frac{8\alpha^{-3}}{(2\pi)^{5/2}} \int_{0}^{\infty} dx \, x^{3/2} e^{(1-s)x} \text{Im} \left[ K_{1/2-iM\alpha}(x) \right].
\end{equation}
By performing the integration and applying the appropriate series expansions for the Macdonald functions, the regularized condensate can be expressed as:
\begin{equation}
    \langle \overline{\psi}\psi \rangle^{(s)} = \frac{1}{\pi^2\alpha^{3}} \text{Im} \left[ \partial_{s}^{2} \sum_{n=0}^{\infty} S_{n} \right],
\end{equation}
where the terms $S_n$ encode the logarithmic and power-law dependence on the regularization parameter $s$:
\begin{equation}
    S_{n} = a_{n} \left[ b_{n} - \ln\left(\frac{s}{2}\right) \right] \left( \frac{s}{2} \right)^{n}.
\end{equation}
The coefficients $a_n$ and $b_n$ are defined as:
\begin{align}
    a_{0} &= 1, \\
    a_{n} &= \frac{iM\alpha (n!)^{-2}}{n+iM\alpha} \prod_{l=1}^{n} (l^{2} + M^{2}\alpha^{2}), \\
    b_{n} &= 2\Psi(n+1) - 2\text{Re}[\Psi(n+iM\alpha)] - \frac{1}{n-iM\alpha},
\end{align}
with $\Psi(z) = \Gamma'(z)/\Gamma(z)$ denoting the digamma function. In the UV limit ($s \to 0$), the $n=0$ term provides the dominant contribution to the gap equation, which, as shown in the subsequent section, leads to the transcendental constraint on the cosmological constant.

\section{Ultraviolet expansion of the fermionic condensate}
\label{sec:UV_expansion}

In this section, we explore the dominant contributions to the fermionic condensate in the ultraviolet (UV) regime. The regularized condensate can be expressed as an asymptotic series in powers of the cutoff parameter $s$, where each term $S_n$ is proportional to $(s/2)^n$~\cite{Saharian:2021mzf}.For small $s$, contributions from $n \ge 3$ are suppressed by higher powers of $s$ and become negligible. Therefore, retaining only the leading-order terms $n = 0, 1, 2$ provides an accurate description of the UV behavior. The regularized condensate is then expressed as:
\begin{equation}
    \langle \overline{\psi}\psi \rangle^{(s)} = \frac{1}{\pi^2\alpha^{3}} \operatorname{Im} \Bigl[ \partial_{s}^{2} \bigl\{ S_{0} + S_{1} + S_{2} \bigr\} \Bigr].
\end{equation}

After performing the derivatives with respect to the regularization parameter $s$ and isolating the imaginary components of the coefficients $a_n$ and $b_n$, we obtain the following expanded form:
\begin{equation}
\label{eq:condensate_full}
\begin{aligned}
\langle \overline{\psi}\psi \rangle^{(s)} &= -\frac{M}{2\pi^{2}\alpha^{2}s} - \frac{M(1+M^{2}\alpha^{2})}{4\pi^{2}\alpha^{2}} \ln\left(\frac{s}{2}\right) \\
&\quad + \frac{M(1+M^{2}\alpha^{2})}{4\pi^{2}\alpha^{2}} \Bigl[ 1 - 2\gamma - 2\operatorname{Re}\Psi(iM\alpha) \Bigr] \\
&\quad - \frac{M}{2\pi^{2}\alpha^{2}},
\end{aligned}
\end{equation}
where $M$ denotes the constituent fermion mass and $\gamma$ is the Euler-Mascheroni constant.

To assign a physical meaning to the regularization parameter $s$, we compare the leading ultraviolet behavior of the regularized de Sitter condensate with the well‑known quadratic divergence in flat spacetime.
From Eq.~\eqref{eq:condensate_full} the dominant term for $s\to0$ is
\begin{equation}
    \langle\overline{\psi}\psi\rangle_{\mathrm{dS}}^{\mathrm{div}} = -\frac{M}{2\pi^{2}\alpha^{2}s},
\end{equation}
whereas in Minkowski space the usual quadratic divergence reads
\begin{equation}
    \langle\overline{\psi}\psi\rangle_{\mathrm{Mink}}^{\mathrm{div}} = -\frac{M\Lambda_{\mathrm{UV}}^{2}}{2\pi^{2}}.
\end{equation}
Equating these two expressions provides a one‑to‑one correspondence between the regularization parameter and the physical ultraviolet cutoff of the effective theory:
\begin{equation}
    s \equiv \frac{1}{\alpha^{2}\Lambda_{\mathrm{UV}}^{2}}.
\end{equation}

This identification ensures that the regularized fermionic condensate in de Sitter space reduces to the standard flat‑space result in the low‑curvature limit $\alpha\to\infty$, and it links the cutoff scale intrinsically to the spacetime geometry. The digamma function $\Psi(iM\alpha)$ encodes curvature effects for arbitrary values of the dimensionless parameter $M\alpha$. To verify the consistency of this formal expansion, we examine two limiting cases.

When the de Sitter radius $\alpha$ is much larger than the fermionic scale, we employ the asymptotic expansion of the digamma function:
\begin{equation}
    \operatorname{Re}\bigl[\Psi(iM\alpha)\bigr] \simeq \ln(M\alpha) + \frac{1}{12(M\alpha)^{2}} + \mathcal{O}\!\left(\frac{1}{\alpha^{4}}\right).
\end{equation}
Substituting this result into the condensate expression and keeping the leading terms, we find:
\begin{equation}
\label{eq:condensate_weak}
\begin{aligned}
\langle \overline{\psi}\psi \rangle^{(s)} &\simeq -\frac{M\Lambda_{\mathrm{UV}}^{2}}{2\pi^{2}} + \frac{M^{3}}{4\pi^{2}} \left\{ 1 - 2\gamma + \ln\left(\frac{2\Lambda_{\mathrm{UV}}^{2}}{M^{2}}\right) \right\} \\
&\quad + \frac{M}{4\pi^{2}\alpha^{2}} \left\{ 1 - 2\gamma + \ln\left(\frac{2\Lambda_{\mathrm{UV}}^{2}}{M^{2}}\right) \right\} - \frac{M}{2\pi^{2}\alpha^{2}}.
\end{aligned}
\end{equation}
In the limit $\alpha \to \infty$, this expression recovers the well-known result for the NJL condensate in flat Minkowski spacetime.

In the regime of strong curvature, $M\alpha \ll 1$, we employ the Taylor expansion of the digamma function near the origin,
\begin{equation}
    \Psi(z) \simeq -\gamma - \frac{1}{z} + \sum_{n=1}^{\infty} (-1)^{n+1}\zeta(n+1)\,z^{n},
\end{equation}
the fermionic condensate simplifies to:
\begin{equation}
\label{eq:condensate_strong}
\begin{aligned}
\langle \overline{\psi}\psi \rangle^{(s)} &\simeq -\frac{M\Lambda_{\mathrm{UV}}^{2}}{2\pi^{2}} + \frac{M}{4\pi^{2}\alpha^{2}} \Bigl[ \ln(2\alpha^{2}\Lambda_{\mathrm{UV}}^{2}) - 1 \Bigr] \\
&\quad + \frac{M^{3}}{2\pi^{2}} \Bigl[ 1 - \zeta(3) + \ln(2\alpha^{2}\Lambda_{\mathrm{UV}}^{2}) \Bigr].
\end{aligned}
\end{equation}
This result serves as the starting point for the exact analytical resolution of the gap equation. As demonstrated in the following section, the dominant logarithmic dependence on $\alpha$ and $\Lambda_{\mathrm{UV}}$ leads to a transcendental constraint on the spacetime geometry mediated by the Lambert $W$ function.

\section{The gap equation and chiral symmetry restoration}
\label{sec:gap_equation}

In this section, we analyze the gap equation in the strong curvature regime ($M\alpha \ll 1$), focusing on the conditions for the restoration of chiral symmetry. Building upon the UV expansion derived in the previous section, the gap equation for a non-vanishing constituent mass $M$ in the presence of a current mass $m$ is given by:
\begin{equation}
\begin{aligned}
\frac{1}{G}-\frac{1}{G_{c}}-\frac{m}{GM} &= -\frac{1}{4\pi^{2}\alpha^{2}}\left[\ln\left(2\alpha^{2}\Lambda_{\text{UV}}^{2}\right)-1\right] \\
&\quad -\frac{M^{2}}{2\pi^{2}}\left[1-\zeta(3)+\ln\left(2\alpha^{2}\Lambda_{\text{UV}}^{2}\right)\right],
\end{aligned}
\end{equation}
where $G_c$ is the critical coupling constant defined in the UV limit as:
\begin{equation}
    \frac{1}{G_{c}} = \frac{\Lambda_{\text{UV}}^{2}}{2\pi^2}.
\end{equation}

We are primarily interested in the dynamical generation of mass and the subsequent restoration of chiral symmetry. In the chiral limit ($m=0$), the gap equation allows us to isolate the constituent mass $M$ as:
\begin{equation}
    M^{2} = \frac{2\pi^{2}\left(\frac{1}{G_{c}}-\frac{1}{G}\right) + \frac{1}{2\alpha^{2}}\left[1-\ln(2\alpha^{2}\Lambda_{\text{UV}}^{2})\right]}{1-\zeta(3)+\ln(2\alpha^{2}\Lambda_{\text{UV}}^{2})}.
\end{equation}
The critical boundary for chiral symmetry restoration is defined by the condition $M=0$. At this threshold, the relation between the coupling constants and the geometric parameters of the de Sitter background simplifies to:
\begin{equation}
    \frac{1}{G} - \frac{1}{G_{c}} = \frac{1}{G_{c}(2\alpha^{2}\Lambda_{\text{UV}}^{2})} \left[ 1 - \ln(2\alpha^{2}\Lambda_{\text{UV}}^{2}) \right].
\end{equation}

This transcendental equation can be solved exactly for the critical de Sitter radius $\alpha_c$. By rearranging the terms into the standard form $ze^z = y$, we obtain the solution in terms of the Lambert $W$ function:
\begin{equation}
    \alpha_{c} = \frac{1}{\sqrt{2}\Lambda_{\text{UV}}} \sqrt{\frac{W_{-1} \left[ -\left(1-\frac{G_{c}}{G}\right)e \right]}{-\left(1-\frac{G_{c}}{G}\right)}},
\end{equation}
where we consider the $W_{-1}$ branch to capture the physically relevant regime of the effective theory. The existence of real solutions for this branch requires the argument of the Lambert function to lie within the domain $[-1/e, 0)$. This mathematical constraint imposes a fundamental inequality:
\begin{equation}
    -e \left( 1 - \frac{G_{c}}{G} \right) \geq -\frac{1}{e}.
\end{equation}

Dividing both sides by $-e$ reverses the inequality, giving 
$1 - G_c/G \leq e^{-2}$, or equivalently $G_c/G \geq 1 - e^{-2}$. 
Taking the reciprocal then yields
a restriction on the coupling ratio, $G/G_{c} \leq (1 - e^{-2})^{-1} \approx 1.15651$.

Furthermore, the branch point at $W_{-1}(-1/e) = -1$ defines the minimum allowable de Sitter radius for the existence of the symmetry-broken vacuum within this effective description:
\begin{equation}
    \alpha_{c}^{\text{min}} = \frac{e}{\sqrt{2}\Lambda_{\text{UV}}}.
\end{equation}
Recalling the relation $\Lambda = 3/\alpha^2$, this result translates into an upper bound for the cosmological constant in terms of the EFT energy scale:
\begin{equation}
    \Lambda \leq \Lambda^{\text{max}} = \frac{6\Lambda_{\text{UV}}^{2}}{e^{2}}.
\end{equation}
Thus, for a given ultraviolet cutoff $\Lambda_{\text{UV}}$, there exists a maximum curvature beyond which the mean-field gap equation no longer admits a real solution for the dynamically generated mass. This indicates a consistency limit of the effective field theory treatment.

\begin{figure}[htbp]
\centering
\includegraphics[width=\linewidth]{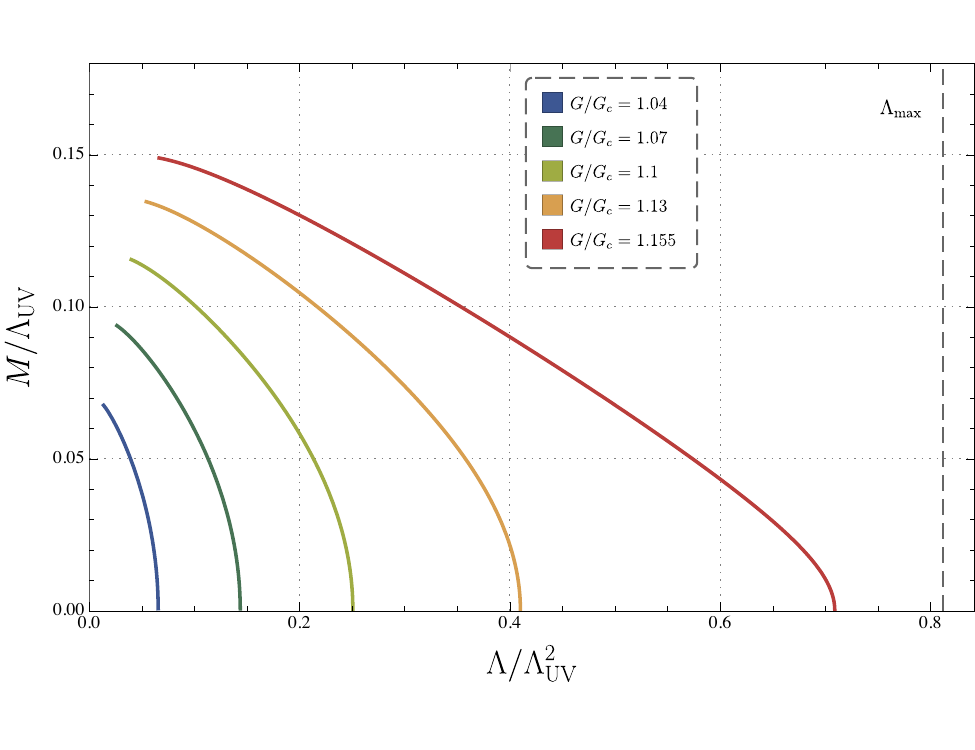}
\caption{Constituent fermion mass $M/\Lambda_\mathrm{UV}$ as a function of the cosmological constant for five values of $G/G_c$. The curves are shown in the region where the solution remains real and terminate at the critical values associated with chiral symmetry restoration.}
\label{fig:mass_lambda}
\end{figure}

Figure~\ref{fig:mass_lambda} shows the constituent fermion mass scaled by $\Lambda_\mathrm{UV}$ as a function of the cosmological constant for five values of $G/G_c$. In this parameter scan, larger couplings keep the mass nonzero up to higher curvature, so the termination point shifts to larger $\Lambda$. The dashed line marks the absolute upper limit $\Lambda^{\max} = 6\Lambda_{\mathrm{UV}}^2/e^2 \approx 0.812\Lambda_{\mathrm{UV}}^2$.

\section{Conclusions}
\label{Conclusions}

We studied chiral symmetry breaking in the Nambu--Jona-Lasinio model on a de Sitter background within an effective field theory framework, treating the ultraviolet cutoff $\Lambda_{\mathrm{UV}}$ as a physical scale. Using exponential proper-time regularization, the fermionic condensate was expanded in the ultraviolet regime and inserted into the gap equation. In the strong-curvature limit, the resulting equation admits an exact analytical solution in terms of the Lambert $W$ function.

The analysis shows that the physically relevant branch, $W_{-1}$, yields real solutions only within a restricted domain. This condition leads to an upper bound on the cosmological constant,
\begin{equation}
    \Lambda^{\max} = \frac{6\Lambda_{\mathrm{UV}}^{2}}{e^{2}},
\end{equation}
as well as a restriction on the coupling ratio, $ G/G_c  \le   1.1565.$ These bounds arise from the consistency of the mean-field solution within the chosen regularization scheme. They should therefore be understood as limits of the effective description, rather than as fundamental constraints on the de Sitter geometry.

The mass curves obtained from the gap equation are consistent with this picture. For larger values of $G/G_c$, the constituent mass remains nonzero up to higher curvature, and the critical point where chiral symmetry is restored is shifted accordingly. The resulting behavior is in agreement with the analytical condition derived from the Lambert $W$ solution.

Possible extensions of this work include the study of higher-dimensional spacetimes and numerical investigations of the gap equation beyond the chiral and strong-curvature regime. Such analyses would help clarify how robust the present results are under changes in the background and in the approximation scheme.

\section{Acknowledgements}\label{Acknowledgements}
L.G.B. acknowledges the financial support of the Coordenação de Aperfeiçoamento de Pessoal de Nível Superior (CAPES), Brazil (Finance Code 001). The author also acknowledges Rafael Pacheco Cardoso and João Victor Zamperlini for fruitful discussions on the Nambu-Jona-Lasinio model.

\bibliographystyle{unsrturl}
\bibliography{sample}

\end{document}